\documentclass{ws-procs9x6}
\begin{document}

\title{RADIATION AND MAGNETIC FIELD EFFECTS ON NEW SEMICONDUCTOR POWER DEVICES FOR HL-LHC EXPERIMENTS}

\author{S. FIORE$^{1,2,*}$, C. ABBATE$^{2,3}$, S. BACCARO$^{1,2}$, G. BUSATTO$^{2,3}$, M. CITTERIO$^{4}$, F. IANNUZZO$^{2,3}$, A. LANZA$^{5}$, S. LATORRE$^{4}$, M. LAZZARONI$^{5,6}$, A.~SANSEVERINO$^{2,3}$ and F. VELARDI$^{2,3}$}

\address{1. UTTMAT, ENEA, via Anguillarese, 301,\\
 Roma, 00123, Italy}

\address{2. INFN Roma, p.le Aldo Moro, 2\\
Roma, 00185, Italy}

\address{3. Dipartimento di Ingegneria Elettrica e dell'Informazione, University of Cassino and Southern Lazio,\\
Cassino, 03043, Italy}

\address{4. INFN Milano, via Celoria 16,\\
Milano, 20133, Italy}

\address{5. INFN Pavia, via Bassi 6,\\
Pavia, 27100, Italy}

\address{6. Dipartimento di Fisica, Universit\`a degli Studi di Milano\\
Milano, 20133, Italy}

\address{$^*$Corresponding author. E-mail: salvatore.fiore@enea.it}

\begin{abstract}
The radiation hardness of commercial Silicon Carbide and Gallium Nitride power MOSFETs is presented in this paper, for Total Ionizing Dose effects and Single Event Effects, under $\gamma$, neutrons, protons and heavy ions. Similar tests are discussed for commercial DC-DC converters, also tested in operation under magnetic field.

\end{abstract}


\bodymatter

\section{The APOLLO collaboration for HL-LHC experiments}\label{aba:intro}
The Phase-2 upgrade of the Large Hadron Collider (LHC) at CERN is planned to take place during year 2022 long shutdown. The present detectors will be upgraded or substituted, and new ones wil be installed to improve the performances and sustain the higher rates and backgrounds at the design luminosity of $5\times 10^{34}$ cm$^{-2}$s$^{-1}$.
 
The increase of radiation background will cause the accumulation of a Total Ionizing Dose (TID) up to 10 kGy (in Silicon), and fluences up to $2\times 10^{13}$ protons/cm$^2$ and $8\times 10^{13}$ neutrons/cm$^2$. This, together with the increased power demand of the detectors, implies a complete re-design of the power distribution system. The new power network will be constitued of isolated main converters, down-stepping the main power voltage of 280 V DC to an intermediate bus at 12 V DC, and Point-Of-Load converters (POL) located at the very heart of the detectors, close to Fornt-End Electronics to be supplied with voltages below 5 V\cite{APOLLO1,APOLLO2}.
In such a position, POL converters will have to be tolerant to the above-mentioned TID and particle fluences, and also to magnetic fields up to 0.1 T generated for the charged particle tracking detectors.

The APOLLO collaboration is testing both single components and devices, commercially available, in order to design Main and POL converters to be operated in the harsh environment of upgraded LHC experiments. 

\section{Silicon Carbide power MOSFETs}
Commercially-available Silicon Carbide (SiC) power MOSFETs  have many advantages from the radiation hardness point-of-view: SiC junction can block a high voltage in volumes much smaller than Silicon, and is also  hard to displacement effects. Minimum energy to displace SiC atoms is in fact larger than in Silicon and GaAs. Single-Event Effect (SEE) sensitivity is then smaller than in equivalent Si power MOSFET.

\begin{figure}
\begin{center}
\psfig{file=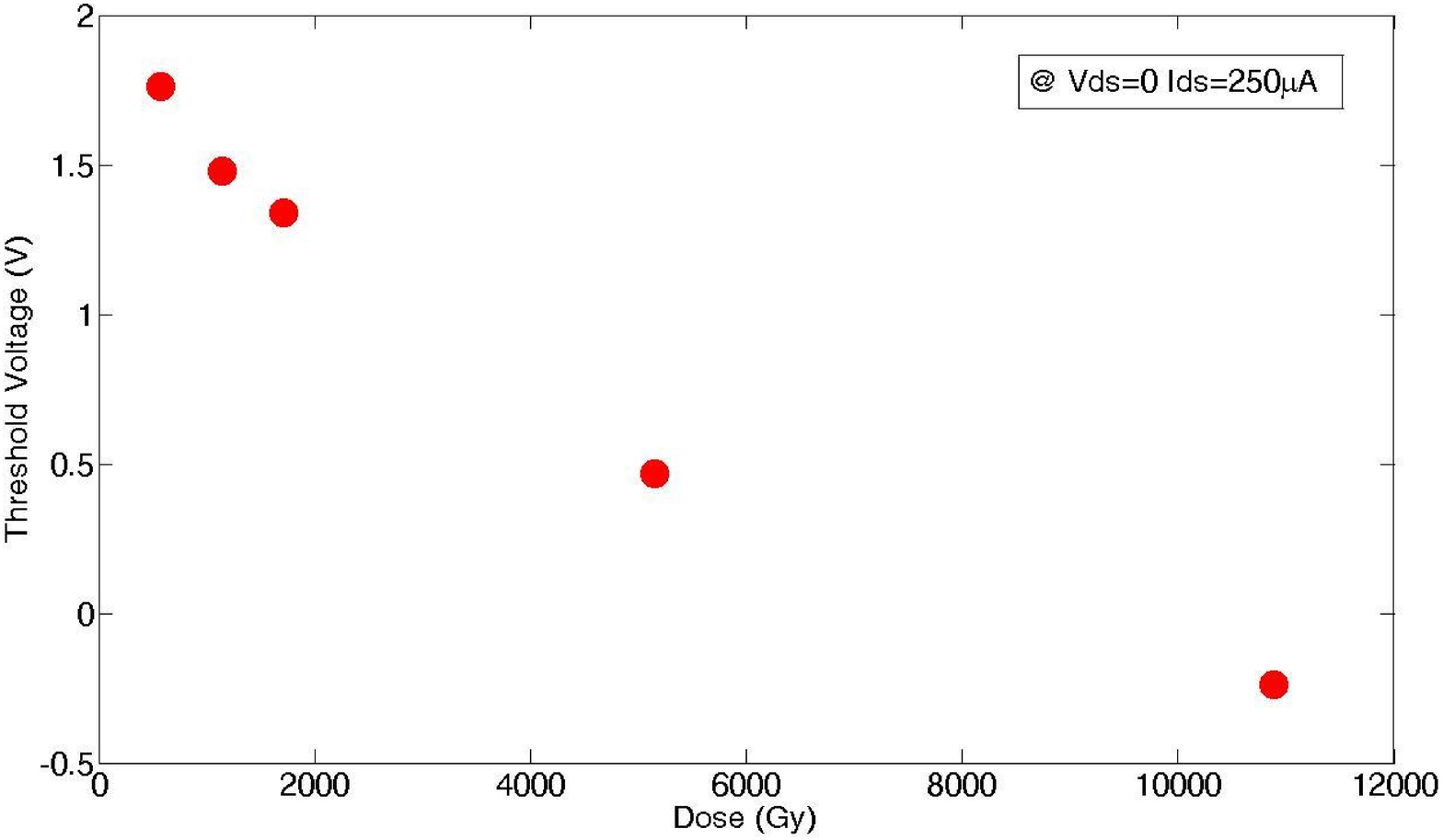,width=2.35in}
\psfig{file=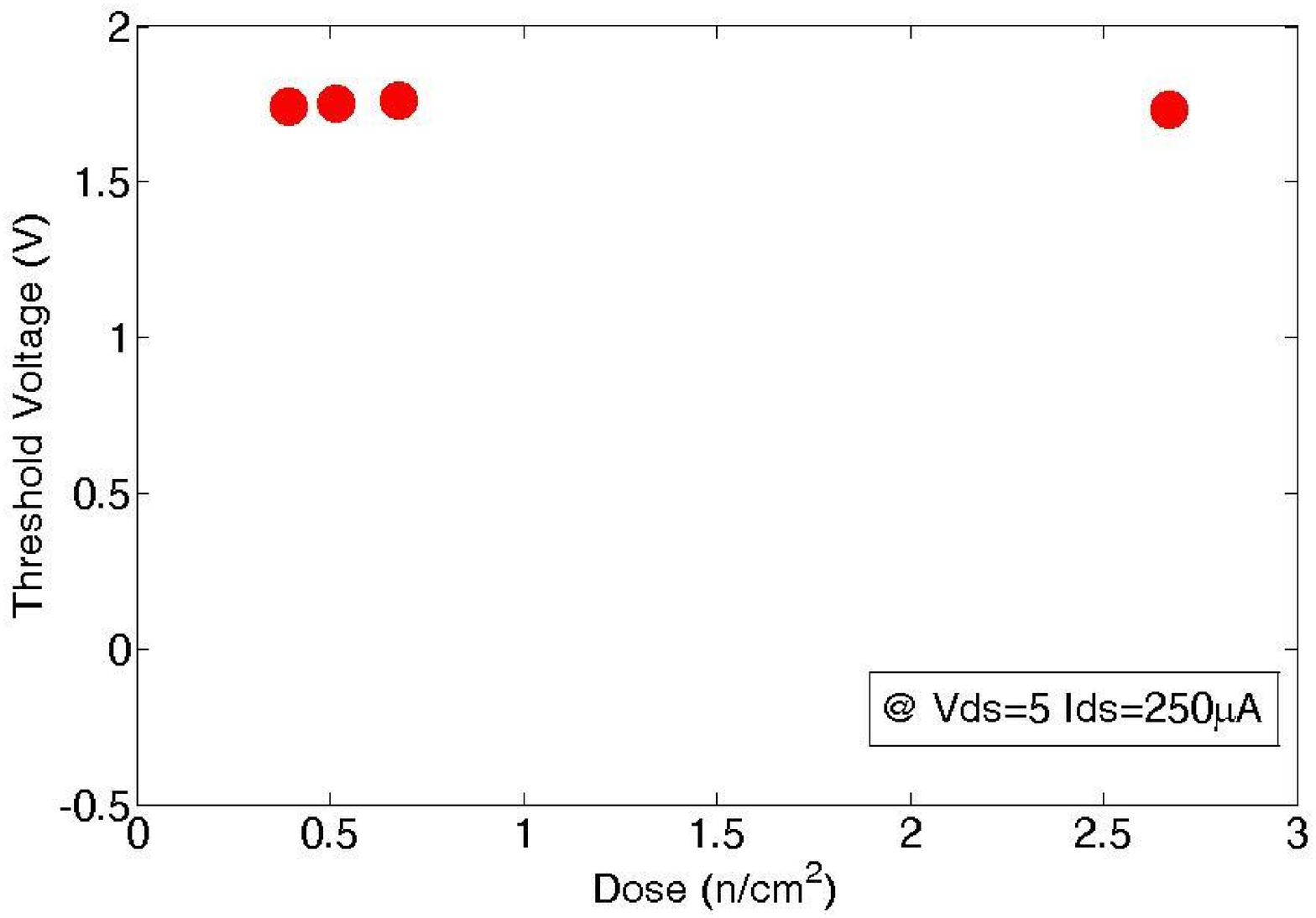,width=2.05in}
\end{center}
\caption{SiC threshold voltage $V_{THR}$ with respect to $\gamma$ TID (left), and to neutron fluence (right).}
\label{aba:fig1}
\end{figure}

SiC devices, manufactured by CREE, have been irradiated with $^{79}$Br ions of 550 MeV, 240 MeV, 60 MeV and 20 MeV energy, in INFN Legnaro Laboratories.
The collected charge due to charge amplification was lower than the one measured in equivalent Si power devices (MOSFETs and IGBTs), due to the smaller SEE sensitive volume. Single-Event Gate Rupture (SEGR) has been detected during $^{79}$Br irradiation, by measuring the Gate current, with 240 MeV beam and 150 V Drain-Source potential (V$_{DS}$), and with 550MeV and V$_{DS}$ = 100 V.
SiC MOSFETS were also exposed to $\gamma$-rays from $^{60}$Co at Calliope Irradiation Facility, and to fast neutrons at Tapiro Research Reactor, both in ENEA Casaccia Research Center. As shown in Fig. (\ref{aba:fig1}), threshold voltage $V_{THR}$ was not affected by 2.7$\times 10^{12}$ neutrons$/cm^2$, while TID of 10.8 kGy (dose rate of 22 Gy/h) induced a drift of -2 V in $V_{THR}$. The latter and the following doses and dose rates by $\gamma$ irradiation are referred to Silicon.

We conclude that SiC MOSFETS are more resistant to Single-Event Burnout (SEB) due to intrinsic SIC properties, but the gate oxide is still vulnerable by TID or by SEGR. This amount of $V_{THR}$ drift is actually not a concern in ATLAS environment, where a negative (-3 V) voltage will be applied to ensure the OFF condition.

\section{Gallium Nitride Enhanced Mode Transistors}
In order to find an alternative to SiC devices, we tested Enhancement-mode Gallium Nitride (GaN) High Electron Mobility Transistors (HEMT) manufactured by EPC.
GaN is quite hard to displacement effects: minimum energy to displace GaN atoms is larger than the one required for Si and GaAs displacement, and is close to SiC one.
In this kind of transistor, the channel is formed through band engineering, so in principle no dielectric layers are used underneath the gate. For this reason, tolerance to total dose is expected to be excellent. Unfortunately no information is provided as to how enhancement-mode has been achieved: additional layers could have been introduced to engineer the positive threshold voltage, and this could have an impact on on radiation hardness.

\begin{figure}
\begin{center}
\psfig{file=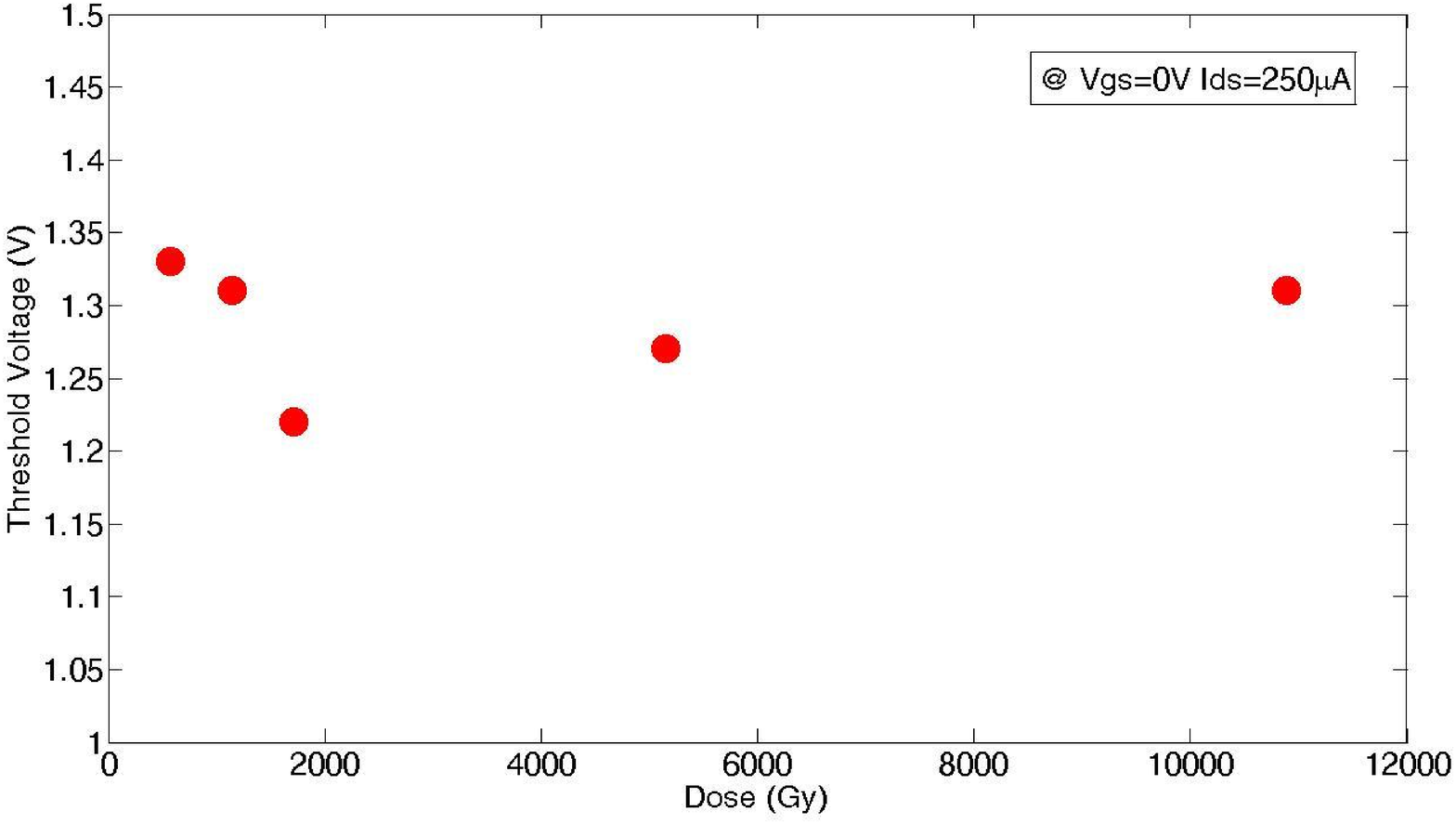,width=2.22in}
\psfig{file=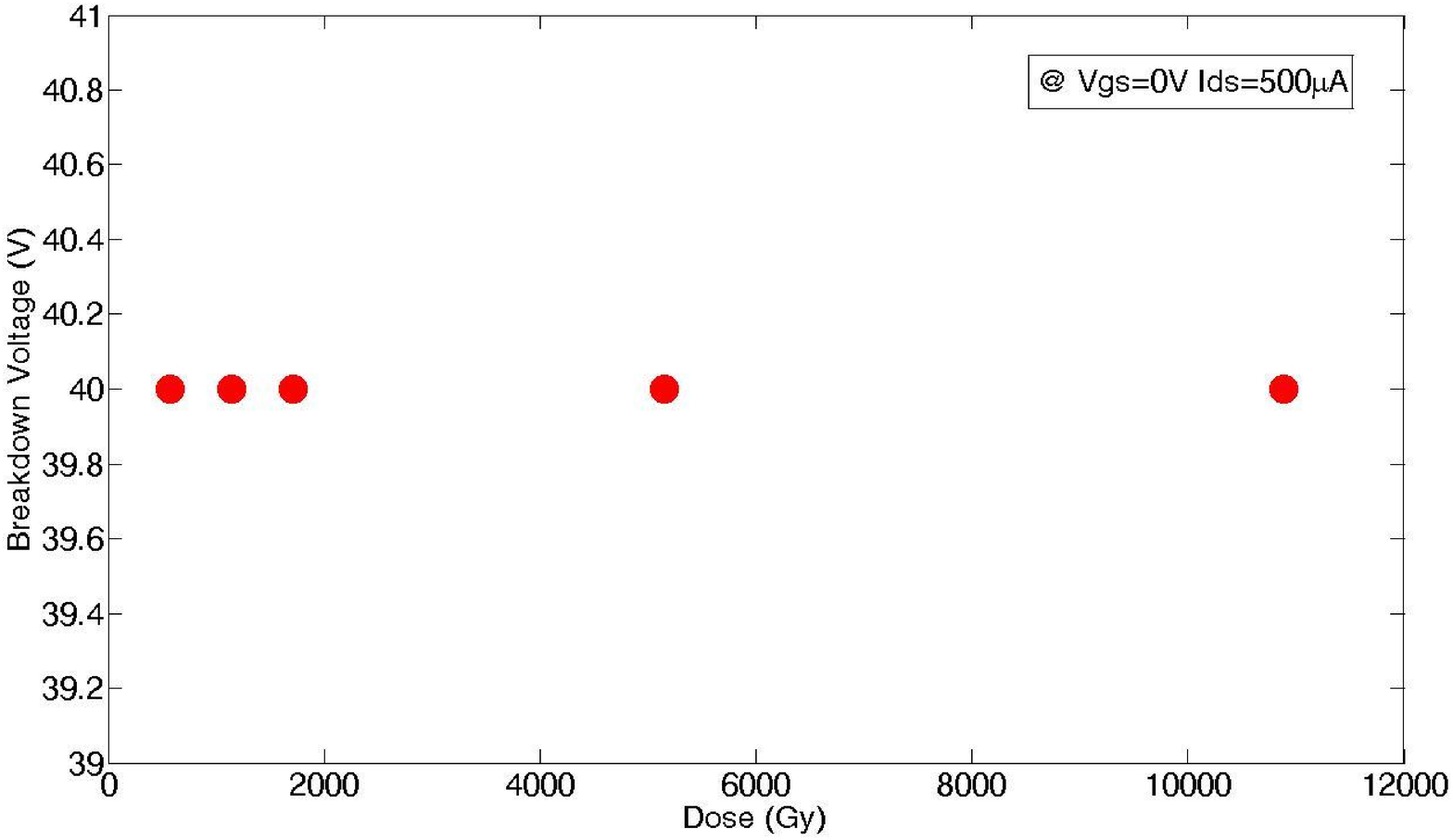,width=2.22in}
\end{center}
\caption{GaN threshold voltage (left) and breakdown voltage (right) with respect to TID, during $\gamma$ irradiation.}
\label{aba:fig2}
\end{figure}

40 V GaN HEMT from EPC were tested with $\gamma$-rays at ENEA Casaccia Calliope facility, with TID up to 10.8 kGy and a dose rate of 22 Gy/h. These devices shown a very good tolerance to TID effects: both threshold voltage and breakdown voltage remain unaffected by the irradiation (Fig. (\ref{aba:fig2})). 

40 V and 200 V GaN HEMT from EPC were also tested with heavy ions (240 MeV $^{79}$Br and 300 MeV $^{125}$I), and no SEE was detected. Under low energy protons irradiation (3 MeV up to 4$\times$10$^{14}$p/cm$^2$) in INFN Legnaro Laboratories, we experienced increase in gate current, threshold voltage reduction and transconductance drop. Since protons induce both ionization and displacement, degradation can be related to both processes, but the total fluence was an order of magnitude higher than what expected in ATLAS phase-2 upgrade.

\section{Linear LTM4619 DC-DC converters}
\begin{figure}
\begin{center}
\psfig{file=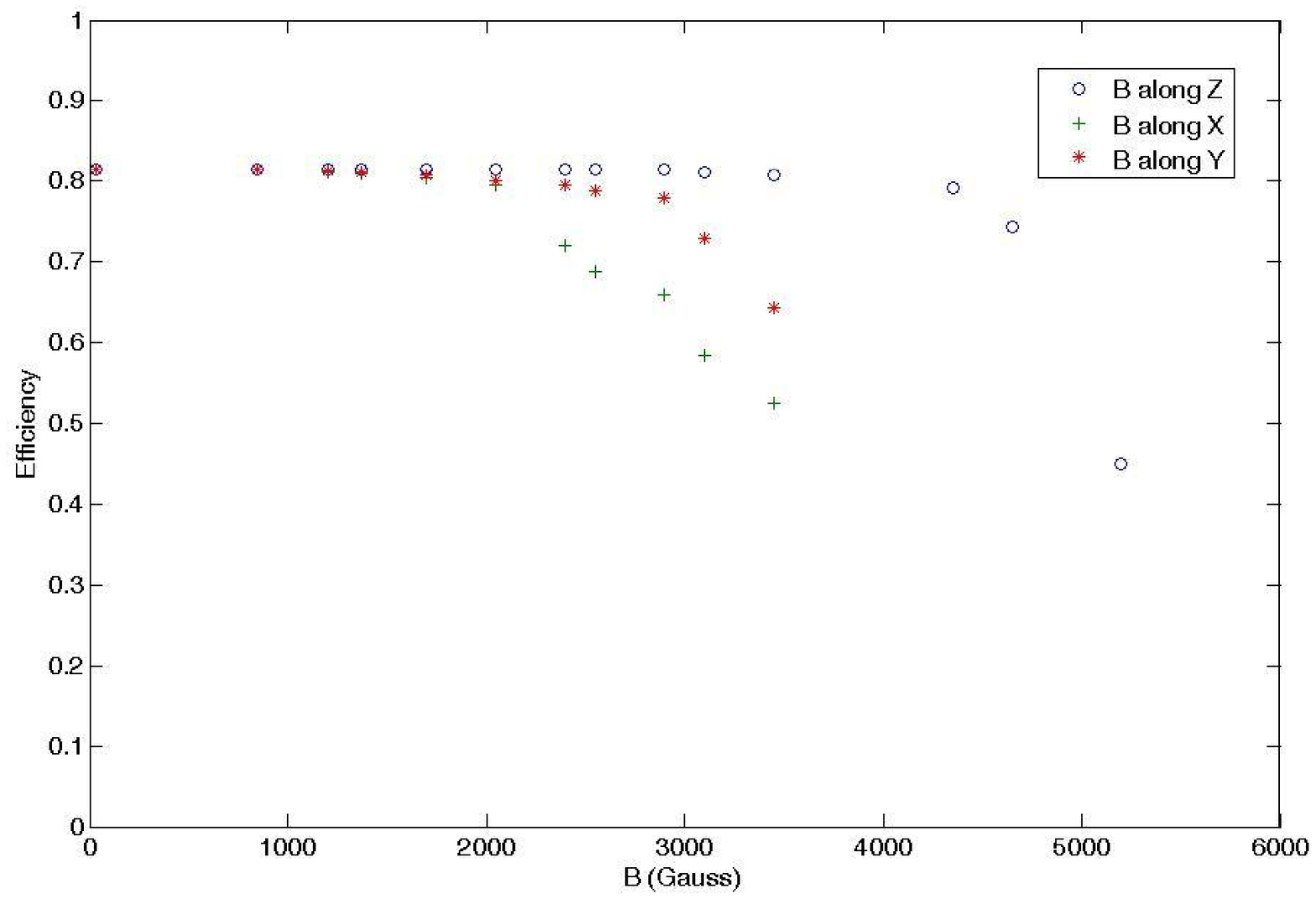,width=2.22in}
\psfig{file=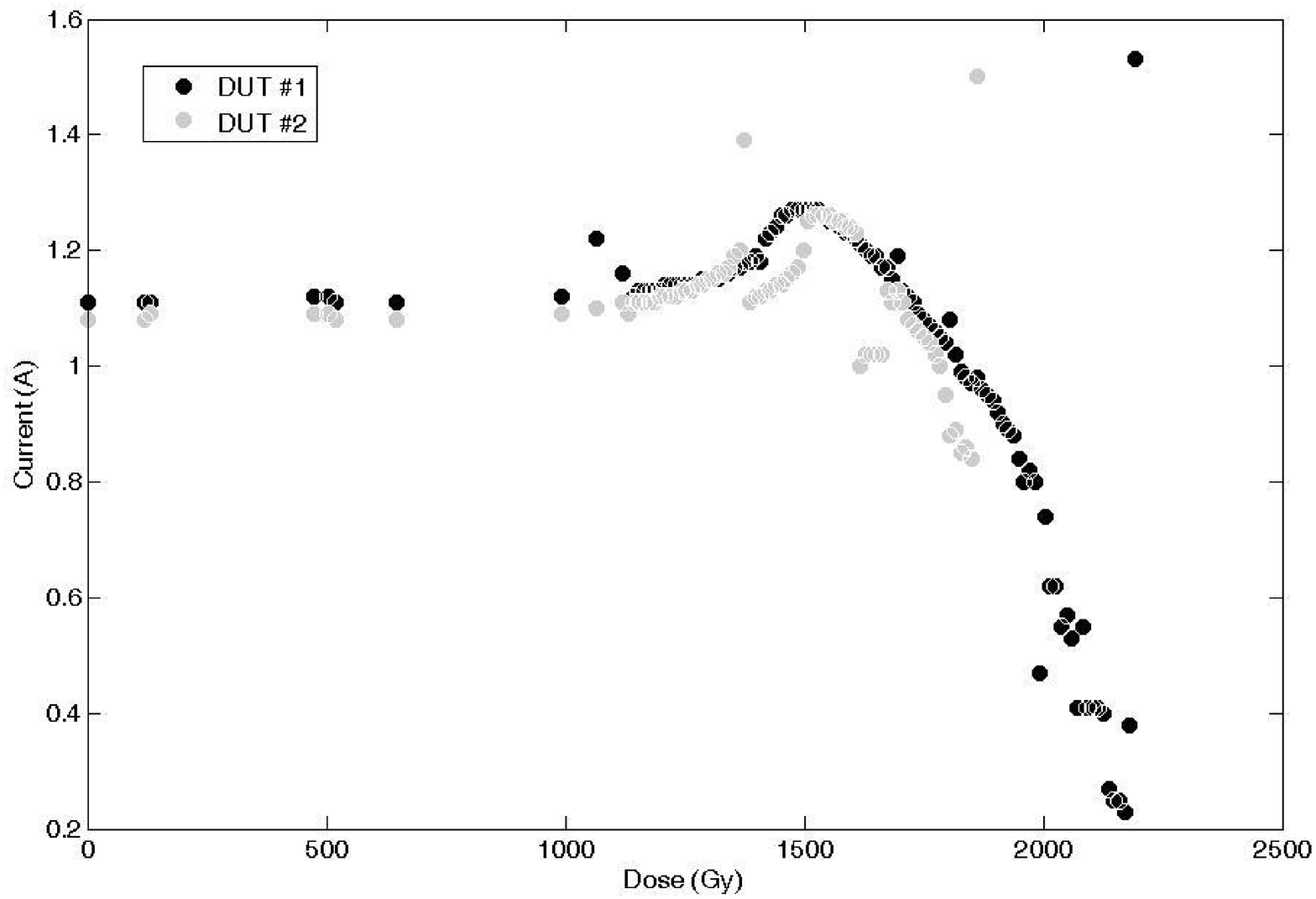,width=2.22in}
\end{center}
\caption{Left: LTM4619 efficiency with respect to B-field, in different orientations. Right: LTM4619 input current with respect to TID, for two devices, during $\gamma$ irradiation.}
\label{aba:fig3}
\end{figure}

An integrated DC-DC step-down converter by Linear Technologies (LTM4619) has been tested for TID and magnetic (B) field tolerance.
To test pourposes we used evaluation boards provided by the manufacturer, with one module mounted per board. Each module has two indipendent outputs and a claimed efficiency of ~80\%.
A test for magnetic field tolerance has been performed in INFN-LASA Laboratory in Milano: one demo board was placed between the polar expansions of a dipole magnet, with B-field aligned along major axes of the converter one at a time. Input voltage was 20 V, both outputs were 1.8 V, 3 A. We evaluated the conversion efficiency $\epsilon = P_{OUT}/P_{IN}$: the efficiency was stable at 83\% up to 0.2--0.4 T B-field, depending on the field orientation, as shown in Fig. (\ref{aba:fig3}).
No permanent effect was observed when the B-field was removed.

Two demo-boards were tested for TID effects at ENEA Calliope $\gamma$ Irradiation Facility, with 2 kGy TID and a dose rate of 22 Gy/h.
Previously, one module has been irradiated at Brookhaven National Laboratories (BNL) by a different research group, but there are no details available about this test.
During our test the outputs were at 1.8 and 3.3 V, 3.4 A for each module, the input was 24 V, the efficiency at test beginning was $\epsilon$ = 68\%, lower than the expected. We monitored the input current to evaluate the change in efficiency.
During the irradiation, the input current increased of some 10\% few times, then returning to initial value. After 1.7 kGy the current dropped continuously until a complete failure of the modules after 2.1 kGy on average. The measurements are shown in Fig. (\ref{aba:fig3}).

Since we could not measure the output voltage, we infer that the current increase was due to an efficiency decrease, while the current drop was the effect of an output voltage decrease due to impaired regulation capabilities.
We will perform TID tests with different dose rates, with a monitoring system for the output voltage. We will also test a different DC-DC converter by Linear (LTM8033 with single output), under TID and B-field.

\section{Conclusions}
SiC power MOSFETs shown good performances in terms of TID under $\gamma$ and neutrons, but exhibited some issues with SEGR in heavy ion irradiation.
EPC Enhancement-mode GaN transistors displayed very good tolerance to TID under $\gamma$ irradiation, and practical immunity from SEEs by heavy ions.
Linear DC-DC converters proved to be insensitive to B-field up to 0.2--0.4 T, and maintained full efficiency up to 1 kGy $\gamma$ TID.



\end{document}